\documentclass{article}
\def\PREPRINT{1}

\ifnum\PREPRINT=0
    \usepackage{spconf}
\else
    \usepackage[preprint]{spconf}
\fi
\usepackage{cite}
\usepackage{amsmath}
\usepackage{amssymb}
\usepackage{amsfonts}
\usepackage{amsthm}
\usepackage{algorithmic}
\usepackage{graphicx}
\usepackage{textcomp}
\usepackage{xcolor}
\usepackage[hidelinks]{hyperref}
\usepackage{pgfplots}
\pgfplotsset{compat=1.18} 
\usepackage[export]{adjustbox}
\usepackage{pifont}
\usepackage{subcaption}
\usepackage{float}
\usepackage{graphicx}
\usepackage{multirow}
\usepackage{bm}
\usepackage{array,booktabs}
\usepackage{balance}
\usepackage{graphicx}
\usepackage{xcolor}
\usepackage{xspace}
\usepackage{balance}

\title{Multi-relational Graph Diffusion Neural Network with Parallel Retention for Stock Trends Classification}
\ifnum\PREPRINT=0
    \name{Zinuo You\textsuperscript{1}, Pengju Zhang\textsuperscript{1}, Jin Zheng\textsuperscript{2}, John Cartlidge\textsuperscript{2}} 

    \address{\textsuperscript{1}School of Computer Science, University of Bristol, UK\\
    \textsuperscript{2}School of Engineering Mathematics and Technology, University of Bristol, UK}
\else 
    \name{Zinuo You\textsuperscript{1}, Pengju Zhang\textsuperscript{1}, Jin Zheng\textsuperscript{2}, John Cartlidge\textsuperscript{2}\thanks{\copyright\ 2024 IEEE. Personal use of this material is permitted.  Permission from IEEE must be obtained for all other uses, in any current or future media, including reprinting/republishing this material for advertising or promotional purposes, creating new collective works, for resale or redistribution to servers or lists, or reuse of any copyrighted component of this work in other works.}} 

    \address{\textsuperscript{1}School of Computer Science, University of Bristol, UK\\
    \textsuperscript{2}School of Engineering Mathematics and Technology, University of Bristol, UK\\
    {\small \tt \{zinuo.you, qo22685, jin.zheng, john.cartlidge\}@bristol.ac.uk}}
\fi

\ninept
\ifnum\PREPRINT=1
   \copyrightnotice{\copyright\ 2024 IEEE.}
   \toappear{To appear in {\it Proc. ICASSP2024, Apr 14-19, 2024, Seoul, Korea.}}
\fi

\begin{document}
\maketitle

\begin{abstract}
Stock trend classification remains a fundamental yet challenging task, owing to the intricate time-evolving dynamics between and within stocks. To tackle these two challenges, we propose a graph-based representation learning approach aimed at predicting the future movements of multiple stocks. Initially, we model the complex time-varying relationships between stocks by generating dynamic multi-relational stock graphs. This is achieved through a novel edge generation algorithm that leverages information entropy and signal energy to quantify the intensity and directionality of inter-stock relations on each trading day. Then, we further refine these initial graphs through a stochastic multi-relational diffusion process, adaptively learning task-optimal edges. Subsequently, we implement a decoupled representation learning scheme with parallel retention to obtain the final graph representation. This strategy better captures the unique temporal features within individual stocks while also capturing the overall structure of the stock graph. Comprehensive experiments conducted on real-world datasets from two US markets (NASDAQ and NYSE) and one Chinese market (Shanghai Stock Exchange: SSE) validate the effectiveness of our method. Our approach consistently outperforms state-of-the-art baselines in forecasting next trading day stock trends across three test periods spanning seven years. Datasets and code have been released.\footnote{\url{https://github.com/pixelhero98/MGDPR}}
\end{abstract}
\section{Introduction}
Predicting the stock market has long been of interest to both academia and industry. In recent years, deep learning techniques have emerged as powerful tools to uncover latent stock market behaviors, particularly in forecasting prices or price movements~\cite{akita2016deep}. Early works typically adopt Recurrent Neural Networks (RNNs) to predict the exact price of a stock based on temporal sequence formed by a single stock indicator such as daily closing prices~\cite{bao2017deep}. While these methods achieved some success in predicting target stock prices, they often overlooked the intricate interconnections between stocks. This can result in catastrophic degradation when 
generalizing to multiple stocks~\cite{deng2019knowledge}. Contemporary approaches bridge the gap by introducing different attention mechanisms~\cite{xu2018stock,yoo2021accurate} or by constructing stock graphs~\cite{kim2019hats,xiang2022temporal}. On the one hand, the attention mechanism considers the relations between stocks by utilizing the element-wise relevance of encoded long temporal sequences. On the other hand, the stock graphs represent stocks as nodes with inter-stock relationships as edges. While the attention-based models~\cite{yoo2021accurate,ding2020hierarchical} demonstrate their efficacy in forecasting stock prices and movements, they have limited ability to model the inter-series relations and capture the temporal features. 
Graph Neural Networks (GNNs) have recently emerged as practical approaches for analyzing complex networks and multivariate time series~\cite{pareja2020evolvegcn,wang2020traffic}. Since the stock market can be regarded as a complex network~\cite{tsay2005analysis}, GNNs offer a well-suited framework for analyzing a diverse set of stocks. Specifically, GNNs explicitly represent individual stocks as nodes, with their time-series data as node attributes, and capture the intricate inter-stock relations through edges.

However, current GNN-based methods for predicting the future status of stocks face two primary challenges: they often neglect the dynamic and asymmetric nature of inter-stock relations~\cite{huynh2023efficient}, and they overlook hierarchical intra-stock features~\cite{sawhney2021exploring}. Traditional GNN-based methods, like HATS~\cite{kim2019hats}, rely on static stock graphs based on industry-corporate or firm-specific relations. Such static stock graphs fail to encapsulate the ever-changing dynamics of stock markets, where inter-stock relations can shift rapidly. Additionally, the hierarchical features intrinsic to individual stocks are frequently sidelined by existing GNN methodologies. Many of these models are built on top of message-passing GNNs, where information propagation and graph representation learning are deeply coupled. This entanglement can result in severe loss of hierarchical intra-stock features~\cite{liu2020towards}, which are crucial for learning both individual and collective stock patterns~\cite{sawhney2021exploring}.

To address the two challenges, we present the Multi-relational Graph Diffusion Neural Network with Parallel Retention (MGDPR), a novel graph-based representation learning approach. The contributions of this work are threefold. First, unlike conventional methods, MGDPR dynamically captures the relations between stocks by leveraging the concepts of information entropy and signal energy. This approach quantifies the directionality and intensity of relationships between stocks at each timestamp, providing a more granular view of the complicated inter-stock dynamics. Moreover, our multi-relational diffusion mechanism refines the generated stock graphs, adaptively learning task-optimal edges that further filter the noisy edges and recover task-relevant edges. Second, we propose a decoupled graph representation learning scheme with parallel retention, preserving the hierarchical intra-stock features and the long-term dependencies of stock indicator time series. Last, through extensive experiments on three real-world datasets covering 2,893 stocks over seven years, we demonstrate the effectiveness of MGDPR in forecasting future stock trends.

\section{Related Work}

\subsection{Attention-based Models for Stock Prediction}

The attention mechanism has achieved great success in natural language processing and time series problems, which stems from its efficacy in handling long sequences and modeling intra-sequence dependencies. Therefore, many attention-based models have been developed for stock selection, price prediction, and movement prediction. For instance, DA-RNN~\cite{qin2017dual} employs a dual-stage attention mechanism within an encoder-decoder framework to forecast future stock prices. This mechanism selectively concentrates on different features of the input time series and further learns hidden inter-series relations. HMG-TF~\cite{ding2020hierarchical} enhances the extraction of hierarchical intra-stock features using a Multi-Scale Gaussian Prior and mitigates redundant learning in multi-head attention through orthogonal regularization. Additionally, DTML~\cite{yoo2021accurate} endeavors to understand the asymmetric and dynamic relations between stocks using a multi-level context transformer in an end-to-end fashion. However, a recurring limitation in these models is their inability to explicitly capture inter-series relations and temporal dynamics, both of which are crucial in multivariate contexts like stock prediction.

\vspace{-0.5em}
\subsection{GNN-based Models for Complex Networks}

In complex evolving networks, entities are often characterized by various time-series features~\cite{liu2019risk}. The stock market, a typical example of such networks, can be effectively modeled by analyzing the inter-relation and intra-features of entities' time series. For instance, HATS~\cite{kim2019hats} integrates Long Short-term Memory and Gated Recurrent Unit layers to distill temporal features as node attributes, further employing a hierarchical attention mechanism to learn graph representations from manually crafted stock graphs. Similarly, HyperStockGAT~\cite{sawhney2021exploring} harnesses hyperbolic graph learning to capture the scale-free nature of inter-stock relations and hierarchical intra-stock features, utilizing a hypergraph attention mechanism to represent spatial stock correlations on Riemannian manifolds. While these studies underscore the importance of retaining hierarchical intra-stock features, their reliance on static, manually crafted stock graphs overlooks the dynamic and asymmetric nature of stock markets. In contrast, GraphWaveNet \cite{wu2019graph} presents an innovative architecture that combines an adaptive dependency matrix with stacked dilated 1D convolution components, adeptly capturing dynamic spatial inter-entity dependencies and managing extended temporal sequences. Furthermore, TPGNN~\cite{liu2022multivariate} captures dynamic inter-entity relations by constructing a matrix polynomial using static matrix bases and time-sensitive coefficients. Nevertheless, these models overlook the complex interplays among various inter-entity relations.

\section{Preliminary}
\label{Preliminary}

\subsection{Notations}
Let $\mathcal{G}_t(\mathcal{V}, \mathcal{E}_{t}, \mathcal{R})$ denote a weighted multi-relational stock graph at trading day $t$, where $\mathcal{V}$ is the set of nodes $\{v_0, ..., v_{N-1}\}$, $\mathcal{E}_{t}$ is the set of edges, and $\mathcal{R}$ is the set of relations. Then, let $\mathbf{A}_{t,r} \in \mathbb{R}^{N \times N}$ denote the adjacency matrix for relation $r \in \mathcal{R}$. Let $\mathbf{X}_{t,r} = \{x_{t,r,0}, x_{t,r,1}, ..., x_{t,r,N-1}\} \in \mathbb{R}^{N \times \tau}$ denote the node feature matrix for relation $r \in \mathcal{R}$, where $\tau$ denotes the historical lookback window size and $x_{t,r,i}$ represents the time series for node $v_i$ under relation $r$. Let $C_t \in \mathbb{R}^{N \times 1}$ denote the label matrix at trading day $t$, and $L$ denotes the number of layers in the model. Let $\mathbf{W}^n_{l}$ denote the learnable weight matrix, $b^n_l$ denote the bias term, and $\sigma(\cdot)$ denotes the activation function.

\subsection{Problem Formulation}
Our approach predicts future trends of multiple stocks by representing them as temporally ordered multi-relational graphs. Hence, we transform this into a node classification task, namely the next trading day stock trend classification, which aligns with previous works~\cite{ding2020hierarchical,sawhney2021exploring}. Meanwhile, we adopt a common yet effective label-generation rule for stocks during training, validation, and test periods. In general, the mapping relationship of this work is defined as,
\begin{equation}
    f(\mathcal{G}_t(\mathcal{V}, \mathcal{E}_{t}, \mathcal{R})) \xrightarrow{} C_{t+\tau_{0}},
    \label{eq1}
\end{equation}
where $f(\cdot)$ denotes the proposed MGDPR and $\tau_0$ denotes the prediction step. In this work, we consider the single-step case with $\tau_0=1$.

\section{Methodology}
\label{Methodology}
\subsection{Dynamic Multi-relational Stock Graph Generation}
Most existing GNN-based approaches for modeling multiple stocks stick to rigid graph generation rules that produce time-invariant stock graphs according to domain or expert knowledge (e.g., industry or firm-specific relations). However, these methods contradict the true nature of the stock market, which changes stochastically over time. According to financial studies~\cite{shahzad2018global,cont2000herd}, the stock market can be modeled as a complicated communication system with stocks as transmitters and receivers of various information. Accordingly, we can further represent the evolving interactions between numerous stocks through a series of stock graphs ordered in time. 

Concretely, we propose a novel multi-relational edge generation method based on information entropy and signal energy. Compared to adopting estimation-complex transfer entropy to obtain the directionality as prior research proposed~\cite{yue2020information}, this allows us to more effectively approximate the directionality and intensity of different relations between stocks at each timestamp. The entry $(a_{t, r})_{i, j}$ of $\mathbf{A}_{t,r}$ is expressed by the following equation,
\begin{align}
    &E(x_{t,r,i}) = \sum_{o=0}^{\tau-1} |x_{t,r,i}[o]|^2, \notag \\
    &H(x_{t,r,i}) = -\sum_{m=0} p(x^{'}_{t,r,i}[m]) \ln{p(x^{'}_{t,r,i}[m])}, \notag \\
    &p(x^{'}_{t,r,i}[m]) = \frac{\sum_{o=0}^{\tau-1} \delta (x^{'}_{t,r,i}[m] - x_{t,r,i}[o])}{\tau}, \notag \\
    &(a_{t, r})_{i, j} = \frac{E(x_{t,r,i})}{E(x_{t,r,j})} e^{H(x_{t,r,i}) - H(x_{t,r,j})},
\end{align}
where $E(\cdot)$ denotes the signal energy, $H(\cdot)$ denotes the information entropy, $x^{'}_{t,r,i}$ denotes the non-repeating subsequence of $x_{t,r,i}$, $p(\cdot)$ denotes the probability of a value, and $\delta(\cdot)$ represents the Dirac delta function. Note that $(a_{t, r})_{i, j}$ represents a directed and weighted edge from $v_i$ to $v_j$, indicating different impacts on each other. This approach facilitates information transmission from nodes with more predictabilities (smaller entropy) to nodes with less predictabilities (larger entropy) while preserving possibilities for the inverse cases.

\subsection{Multi-relational Graph Diffusion}
Conventional message-passing GNNs, such as Graph Convolutional Neural Networks (GCNs)~\cite{kipf2016semi} and Graph Attention Neural Networks (GATs)~\cite{velivckovic2017graph}, propagate information by aggregating features from immediate neighbors of nodes. Nonetheless, this may overlook the potential incompatibility between the provided graph and the task objective, where certain edges are task-irrelevant and lead to sub-optimal performance~\cite{chen2020measuring}. To alleviate this issue, we propose a multi-relational diffusion process to spread different types of information through learned task-optimal edges adaptively. The multi-relational graph diffusion at layer $l$ is defined as,
\begin{align}
    &\mathbf{S}_{l, r} = (\sum_{k=0}^{K-1}\gamma_{l,r,k}\mathbf{T}_{l, r, k})\odot\mathbf{A}_{t,r},\;\sum_{k=0}\gamma_{l,r,k} = 1, \label{eq2}\notag\\&
    \mathbf{H}_l = \sigma\left(\text{Conv2d}_{1\times1}\left(\Delta(\mathbf{S}_{l, r}\mathbf{H}_{l-1}\mathbf{W}^r_l)^{|\mathcal{R}|-1}_{r=0}\right)\right), 
\end{align}
where $\gamma_{l,r,k}$ represents the weight coefficient, $\mathbf{T}_{l,r,k}$ denotes the column-stochastic transition matrix, $\mathbf{S}_{l,r}$ denotes the diffusion matrix, $K$ denotes the expansion step, $\text{Conv2d}_{1\times1}(\cdot)$ denotes the 2D convolutional layer with a 1x1 kernel, $\Delta(\cdot)^{|\mathcal{R}|-1}_{r=0}$ denotes the function that stacks all relational node representations, and $\mathbf{H}_{l}$ represents the latent diffusion representation.

The weight coefficient should satisfy the constraint to prevent graph signals from being amplified or reduced during the multi-relational graph diffusion. As the relations of graphs are temporally evolving in this work, we make $\gamma_{l,r,k}$ and $\mathbf{T}_{l,r,k}$ as learnable parameters rather than fixing on the invariant mappings for fixed-relation graphs (e.g., CORA, PubMed, CiteSeer, etc.) of the original graph diffusion~\cite{gasteiger2019diffusion}. 

\begin{table}[tb]
\renewcommand\arraystretch{1.25}
    \centering
    \caption{Dataset description for the three markets.}
    \vspace{-0.5em}
    \resizebox{\linewidth}{!}{%
    \begin{tabular}{c| c c c}
    \hline
        & NASDAQ & NYSE & SSE\\
        \hline
        Training Period (Tr) & 01/2013-12/2014 & 01/2013-12/2014 & 01/2015-12/2016\\
        Validation Period (Val) & 01/2015-06/2015 & 01/2015-06/2015 & 01/2017-12/2017\\
        Test Period (Test) & 07/2015-12/2017 & 07/2015-12/2017 & 01/2018-12/2019\\
        \# Days (Tr:Val:Test) & 483:103:631 & 468:123:631 & 488:241:503\\
        \# Stocks & 1026 & 1737 & 130\\
        \# Stock Indicators ($|\mathcal{R}|$) & 5 & 5 & 5\\
        \hline       
    \end{tabular}%
    }
    \label{tabdata}
\end{table}

\subsection{Graph Representation Learning with Parallel Retention}
From the perspective of the expressiveness of GNNs, the deeply intertwined process of information propagation and representation learning hinders learning unique information of individual nodes~\cite{zeng2021decoupling}. This can result in a less expressive learned graph representation, degrading the model performance~\cite{liu2020towards}. Prior research, such as DAGNN~\cite{liu2020towards}, instead of directly utilizing the representation from the final GNN layer, adopts an attention mechanism to learn a weighted graph representation based on representations from different layers. Similarly, approaches like HyperStockGAT~\cite{sawhney2021exploring} learn the graph representation in the hyperbolic space with attention mechanisms. This preserves distinctive hierarchical features of stock graphs, leading to substantial model improvements. On the one hand, the attention mechanism primarily concentrates on the relevance between different segments of the input sequence but falls short in modeling long-term dependencies. On the other hand, the retention mechanism is designed to capture the long-term dependencies while retaining the ability to model the content-wise relevance~\cite{sun2023retentive}. 

\begin{figure}[tb]
    \centering
    \includegraphics[scale=0.5]{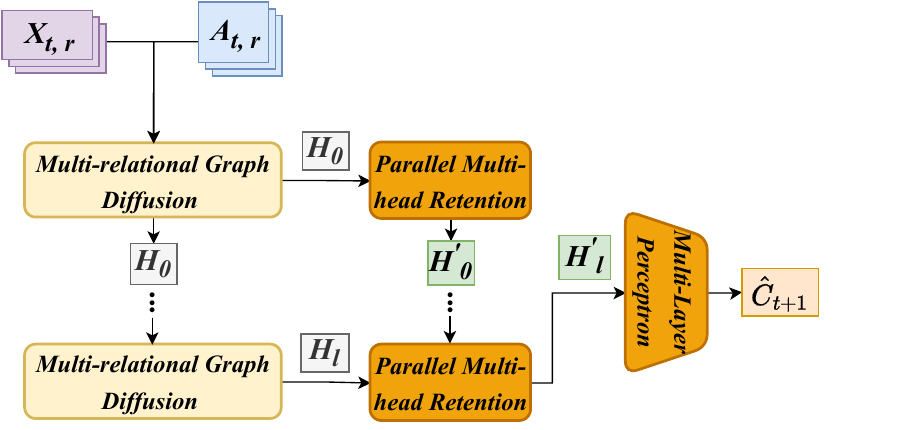}
    \caption{Schematic of graph representation learning scheme.}
    \label{arc}
\end{figure}

Consequently, we propose replacing the attention mechanism with a parallel retention mechanism to better incorporate long-term dependencies in the stock time series. The overall architecture of the proposed MGDPR is presented in Fig.~\ref{arc}. The layer update rule of MGDPR is defined by,
\begin{equation}
\mathbf{H}_{l}^{'} =  \sigma\left((\eta\left(\mathbf{H}_l\right)||(\mathbf{H}^{'}_{l-1}\mathbf{W}^1_{l}+b^1_l))\mathbf{W}^2_l + b^2_l\right)
\end{equation}
where $\mathbf{H}_{l}^{'}$ denotes the hidden graph representation, $||$ denotes concatenation, and $\eta(\cdot)$ denotes the parallel retention.\footnote{We have compared a variant of the model with multi-head attention \cite{zinuo2023decoupled}. The model with parallel retention has slightly higher performance and also benefits from computational efficiencies in training.} 

Parallel retention is defined by,
\begin{align}
    &\mathbf{Q} = \mathbf{Z}\mathbf{W}_{Q}, \quad \mathbf{K} = \mathbf{Z}\mathbf{W}_{K}, \quad \mathbf{V} = \mathbf{Z}\mathbf{W}_{V} \notag\\
    &\mathbf{D}_{ij} =  \left\{
        \begin{array}{ll}
            \zeta^{i-j}, & \text{if } i \geq j \\
            0, & \text{if } i \leq j 
        \end{array}
    \right. \notag\\
    &\eta(\mathbf{Z}) = \phi\left((\mathbf{Q}\mathbf{K}^{T} \odot \mathbf{D})\mathbf{V}\right).
\end{align}
Here, $\phi(\cdot)$ denotes the group normalization, $\mathbf{D}_{ij}$ denotes the entry of $\mathbf{D} \in \mathbb{R}^{\tau \times \tau}$ which is the masking matrix composing causal masking and decay factors along the relative distance, and $\zeta$ denotes the decay coefficient.

\begin{table*}[tb]
\renewcommand\arraystretch{1.25}
\caption{Test period evaluation for MGDPR and baselines. Higher values indicate better performance for all metrics; bold denotes best.}%
\vspace{-0.5em}
\centering
\resizebox{\linewidth}{!}{%
\begin{tabular}{c| c c c| c c c| c c c}
  \hline
   \multirow{2}{*}{Method} & \multicolumn{3}{c|}{NASDAQ} & \multicolumn{3}{c|}{NYSE} & \multicolumn{3}{c}{SSE}\\

   & Acc(\%) & Mcc & F1  & Acc(\%)  & Mcc & F1 & Acc(\%)  & Mcc & F1\\ 
\hline
  DA-RNN~\cite{qin2017dual} & 56.33$\pm$1.15 & 0.04$\pm$4.06$\times10^{-3}$ & 0.54$\pm$0.02 & 57.28$\pm$0.76 & 0.05$\pm$2.23$\times10^{-3}$ & 0.56$\pm$0.01 & 57.03$\pm$0.42 & 0.04$\pm$2.41$\times10^{-3}$ & 0.55$\pm$0.02 \\
  HMG-TF~\cite{ding2020hierarchical} & 58.32$\pm$0.41 & 0.10$\pm$1.79$\times10^{-3}$ & 0.57$\pm$0.01 & 59.11$\pm$0.35 & 0.09$\pm$3.81$\times10^{-3}$ & 0.59$\pm$0.01 & 58.90$\pm$0.36 & 0.11$\pm$3.01$\times10^{-3}$ & 0.59$\pm$0.01 \\
  DTML~\cite{yoo2021accurate} & 57.56$\pm$0.67 & 0.06$\pm$1.98$\times10^{-3}$ & 0.58$\pm$0.01 & 58.78$\pm$0.45 & 0.08$\pm$2.00$\times10^{-3}$ & 0.60$\pm$0.01 & 59.63$\pm$0.21 & 0.09$\pm$5.42$\times10^{-3}$ & 0.59$\pm$0.01\\
  HATS~\cite{kim2019hats} & 50.37$\pm$1.80 & 0.01$\pm$4.79$\times10^{-3}$ & 0.48$\pm$0.02 & 51.93$\pm$0.76 & 0.02$\pm$6.55$\times10^{-3}$ & 0.50$\pm$0.03 & 53.13$\pm$0.47 & 0.02$\pm$5.07$\times10^{-3}$ & 0.50$\pm$0.01\\
  GraphWaveNet~\cite{wu2019graph} & 59.19$\pm$0.55  & 0.06$\pm$6.83$\times10^{-3}$ & 0.60$\pm$0.02 & 62.14$\pm$1.08 & 0.07$\pm$3.20$\times10^{-3}$ & 0.59$\pm$0.02 & 60.78$\pm$0.23 & 0.06$\pm$2.93$\times10^{-3}$ & 0.57$\pm$0.01 \\
  HyperStockGAT~\cite{sawhney2021exploring} & 57.23$\pm$0.71  & 0.06$\pm$5.36$\times10^{-3}$ & 0.59$\pm$0.02  & 59.34$\pm$0.46  & 0.08$\pm$5.73$\times10^{-3}$ & 0.61$\pm$0.02 & 58.36$\pm$0.22  & 0.09$\pm$4.10$\times10^{-3}$ & 0.58$\pm$0.02 \\
  TPGNN~\cite{liu2022multivariate} & 60.42$\pm$0.49  & 0.10$\pm$3.45$\times10^{-3}$ & 0.61$\pm$0.02  & 61.81$\pm$0.19 & 0.11$\pm$4.45$\times10^{-3}$ & 0.60$\pm$0.02 & 62.69$\pm$0.10  & 0.12$\pm$1.66$\times10^{-3}$ & 0.60$\pm$0.02 \\
  \hline
  MGDPR & \textbf{62.77$\pm$0.65}  & \textbf{0.13$\pm$4.49$\times10^{-3}$} & \textbf{0.62$\pm$0.01}  & \textbf{64.54$\pm$0.20}  & \textbf{0.13$\pm$1.88$\times10^{-3}$} & \textbf{0.63$\pm$0.01} & \textbf{63.90$\pm$0.32}  & \textbf{0.14$\pm$2.01$\times10^{-3}$} & \textbf{0.62$\pm$0.02}  \\
  \hline
  \end{tabular}%
  }
\label{tab11}
\end{table*}

\nocite{zinuo2023decoupled}

\subsection{Objective Function}
Considering Eq.~\ref{eq2} and Eq.~\ref{eq1}, the model is optimized towards minimizing the objective function as defined,
\begin{align}
\mathcal{J}&=\frac{1}{B}\sum_{t=0}^{B-1}\mathcal{L_{CE}}(\hat{C}_{t+1}, C_{t+1}) + 
    \sum_{l=0}^{L-1}\sum_{r=0}^{|\mathcal{R}|-1}(\sum_{k=0}^{K-1}\gamma_{l,r,k} - 1).
\end{align}
Here, $\hat{C}_{t+1}$ denotes the predicted label matrix of the trading day $t+1$, $\mathcal{L}_{CE}(\cdot)$ denotes the cross-entropy loss, the second term denotes the multi-relational graph diffusion constraint for all relations and all layers, and $B$ denotes the batch size.

\section{Experiments}
\label{Experiments}
\subsection{Experiment Setup}
All methods were implemented by Pytorch, Cuda Version 12.0, with 2x Navida A100.

\noindent
\textbf{Dataset}. Following previous works, the stock data (active on 98\% trading days) used in this work are collected from two US stock markets (NASDAQ and NYSE) and one China stock market (SSE). The stock indicators of NASDAQ, NYSE, and SSE are open price, high price, low price, close price, and trading volume. The details of the three datasets are presented in Table~\ref{tabdata}.

\noindent
\textbf{Evaluation Metric}. Following prior research on stock classification and stock prediction, we use three metrics to evaluate the performance of the model on the downstream task, classification accuracy (Acc), Matthews correlation coefficient (Mcc), and F1-Score (F1).

\noindent
\textbf{Model Setting}. The historical lookback window size $\tau$ is set to 21, which coincides with values from previous works and professional financial studies~\cite{adam2016stock}. The number of layers $L$ for NASDAQ and NYSE is set to 8, and for SSE is set to 5. The embedding dimension is set to 256. The number of layers of MLP is set to 2 for NASDAQ, NYSE, and SSE. The batch size is set to full batch, the learning rate is set to $2.5\times 10^{-4}$, the number of training epochs is set to 900, and the optimizer is set to Adam. The decay coefficient $\zeta$ is set to 1.27, and the expansion step $K$ is set to 7 for NASDAQ, 8 for NYSE, and 3 for SSE. All hyperparameters are tuned on the validation periods.

\noindent
\textbf{Baseline}. We further compare our approach with methods related to stock movements prediction, including attention-based methods: DA-RNN~\cite{qin2017dual}, DTML~\cite{yoo2021accurate}, and HMG-TF~\cite{ding2020hierarchical}, and GNN-based methods: HATS~\cite{kim2019hats}, GraphWaveNet~\cite{wu2019graph}, HyperStockGAT~\cite{sawhney2021exploring}, and TPGNN~\cite{liu2022multivariate}.

\vspace{-0.5em}
\subsection{Experimental Result and Analysis}

We evaluate the performance of our proposed model MGDPR against other baseline methods on the next trading day stock trend classification across three test periods. From the results summarized in Table~\ref{tab11}, we can make three observations. (i) Graph-based methods like GraphWaveNet, TPGNN, MGDPR, and HyperStockGAT, which explicitly model the relationships between entities, outperform those such as DA-RNN, HMG-TF, and DTML, which rely on attention mechanisms for implicit inter-entity relations. (ii) Among GNN-based models, those (TPGNN, MGDPR, and GraphWaveNet) capturing dynamic links and broader neighborhoods yield superior results compared to HyperStockGAT and HATS. This advantage may stem from the latter methods' reliance on pre-defined static stock graphs and updates limited to immediate neighbors. (iii) Among models incorporating dynamic relations, MGDPR performs better over TPGNN and GraphWaveNet, which only consider single inter-entity relations. In summary, the proposed MGDPR consistently outperforms other models across all evaluation metrics. These observations indicate that neither prior domain knowledge nor manually constructed graphs adequately capture the complex, evolving relationships between stocks. Previous studies have relied on undirected and unweighted graphs, which are insufficient for modeling the dynamic nature of stock markets. In the meantime, the downstream task involves labeling multiple stocks at each timestamp, necessitating models that incorporate the likelihood that highly similar stocks may belong to different classes or seemingly unrelated stocks may fall into the same class over time. Therefore, it is crucial to consider the time-series features of individual stocks in terms of both piece-wise relevance and long-term dependencies.

\vspace{-0.5em}
\subsection{Ablation Study}

\begin{figure}[ht]
\centering
\includegraphics[width=\linewidth]{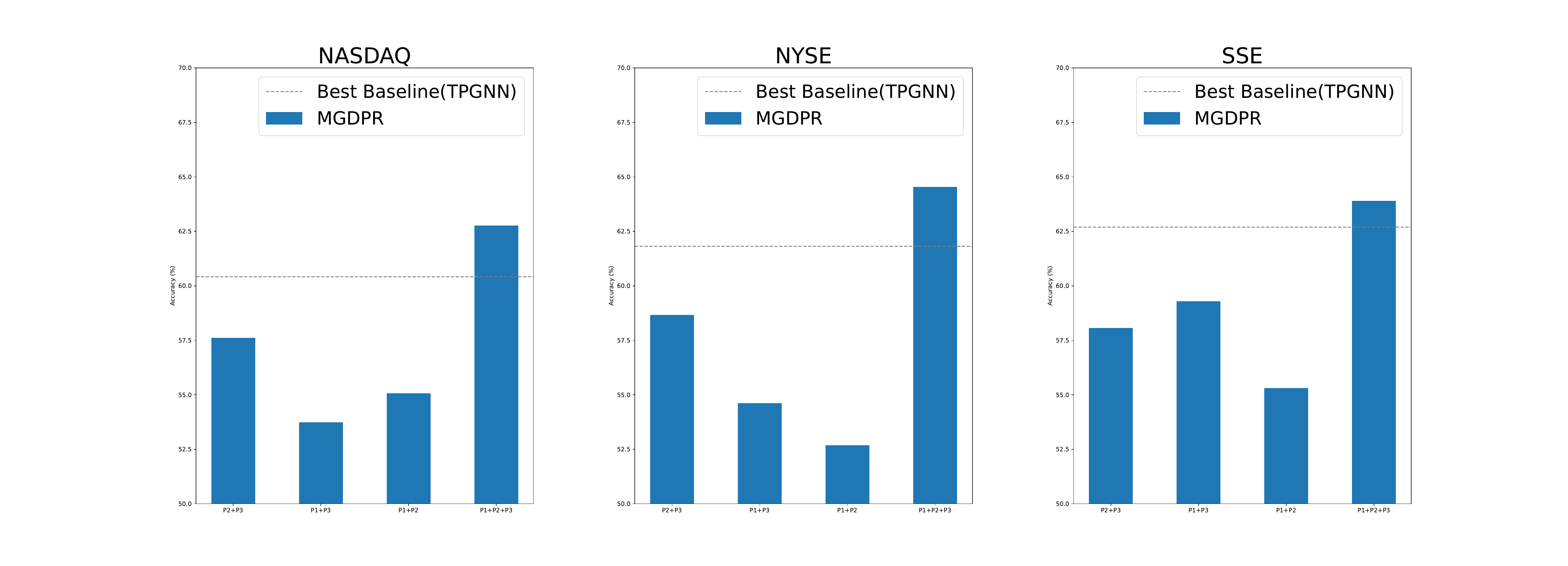}
\caption{Ablation study results. Only the full MGDPR (P1+P2+P3) has higher accuracy than the best baseline in each market.}
\label{fig:abl}
\end{figure}

We conduct an ablation study to investigate the effectiveness of each module in our proposed approach, shown in Fig.~\ref{fig:abl}. The proposed MGDPR incorporates three critical components: Dynamic Multi-relational Stock Graph Generation (P1), Multi-relational Graph Diffusion (P2), and Graph Representation Learning with Parallel Retention (P3). Specifically, removing P1 leads to an average decrease in classification accuracy by 3.42\%, underscoring the importance of dynamically capturing multiple inter-stock relations. Eliminating P2 has varying impacts across the three datasets, likely due to differences in graph sizes (i.e., NASDAQ and NYSE have more stocks than SSE). The diffusion process proves more effective in larger stock graphs, possibly as it adaptively disseminates information across broader neighborhoods. Lastly, replacing P3 results in an average decline in classification accuracy by 6.32\%. This suggests that the parallel retention mechanism better captures long-term dependencies and temporal features of stock time series.

\section{Conclusion}
We have presented MGDPR, a novel graph learning framework designed to capture evolving inter-stock relationships and intra-stock features. To address the limitations of traditional graph-based models,  we conceptualize the stock market as a communication system. First, we use information entropy and signal energy to quantify the connectivity and intensity between stocks on each trading day. Then, we propose the multi-relational diffusion process for the generated stock graph, aiming to learn task-optimal edges. This step alleviates the discordance between the task objectives and the input graph. Finally, we adopt a decoupled graph representation learning scheme with parallel retention. This module is designed to preserve hierarchical features within individual stocks and the long-term dependencies in their time-series features. Empirical results validate the efficacy of MGDPR in forecasting future stock trends. One limitation is MGDPR's focus on quantitative representations of information entropy, while neglecting statistical properties. Future work will optimize the entropy-based algorithm through statistical physics theories.

\ifnum\PREPRINT=0
    \vfill\pagebreak
    \clearpage
\fi
\balance

\bibliographystyle{IEEEbib}

\ifnum\PREPRINT=0
    \bibliography{strings,refs}
\else
    {\footnotesize \bibliography{strings,refs}} 
\fi

\end{document}